\documentstyle[prl,aps,psfig,floats,preprint]{revtex}

\def\beq{\begin{equation}}
\def\eeq{\end{equation}}
\def\ber{\begin{eqnarray}}
\def\eer{\end{eqnarray}}

\def\lsim{\
  \lower-1.5pt\vbox{\hbox{\rlap{$<$}\lower5.3pt\vbox{\hbox{$\sim$}}}}\ }
\def\gsim{\
  \lower-1.5pt\vbox{\hbox{\rlap{$>$}\lower5.3pt\vbox{\hbox{$\sim$}}}}\ }

\begin{document}
\tightenlines

\title{Bouncing Braneworlds}

\author{Yuri Shtanov$^a$ and Varun Sahni$^b$}
\address{$^a$Bogolyubov Institute for Theoretical Physics, Kiev 03143, Ukraine\\
$^b$Inter-University Centre for Astronomy and Astrophysics, Post Bag 4,
Ganeshkhind, Pune 411~007, India}
\maketitle

\begin{abstract}
We study cosmological braneworld models with a single {\em timelike\/} extra
dimension. Such models admit the intriguing possibility that a contracting
braneworld experiences a natural bounce without ever reaching a singular state.
This feature persists in the case of anisotropic braneworlds under some
additional  and not very restrictive assumptions. Generalizing our study to
braneworld models containing an induced brane curvature term, we find that a
FRW-type singularity is once again absent if the bulk extra dimension is
timelike. In this case, the universe either has a non-singular origin or
commences its expansion from a quasi-singular state during which both the
Hubble parameter and the energy density and pressure remain finite while the
curvature tensor diverges. The non-singular and quasi-singular behaviour which
we have discovered differs both qualitatively and quantitatively from what is
usually observed in braneworld models with spacelike extra dimensions and could
have interesting cosmological implications.
\end{abstract}


Following the seminal papers \cite{ADD,RS}, most braneworld models with extra
dimensions assume that extra dimensions are spacelike, so that the brane is
embedded in a Lorentzian multidimensional manifold.  However, there is no {\it
a~priori\/} reason why extra dimensions cannot be {\it timelike\/}, and the
observational constraints on such braneworld models were discussed in
\cite{DGS}.  More recently, such models were under consideration in
\cite{CK,BCKY,Kofinas,Shtanov,SS}. In this letter, we demonstrate that a
timelike extra dimension could, in the case of the simplest braneworld model,
lead to interesting new features. For instance, a contracting braneworld
generically bounces as it reaches a high density thereby leading to the absence
of the `big crunch' and `big bang' singularities of general relativity. We
consider both Friedmann--Robertson--Walker (FRW) and Bianchi~I scenarios and
demonstrate that, under some additional and not very restrictive assumptions,
the presence of anisotropy does not modify this conclusion.

In this letter, we consider the case where a four-dimensional hypersurface
(brane) is the boundary of a five-dimensional Riemannian manifold (bulk) with
nondegenerate Lorentzian induced metric. The action of the theory has the
natural general form
\begin{equation} \label{action}
S = M^3  \left[ \int_{\mathrm bulk} \left({\mathcal R} - 2 \Lambda \right) - 2
\epsilon \int_{\mathrm brane} K  \right] + \int_{\mathrm brane} \left( m^2 R -
2 \sigma \right) + \int_{\mathrm brane} L (h_{ab}, \phi) \, .
\end{equation}
Here, ${\mathcal R}$ is the scalar curvature of the five-dimensional metric
$g_{ab}$ in the bulk, and $R$ is the scalar curvature of the induced metric
$h_{ab} = g_{ab} - \epsilon n_a n_b$ on the brane, where $n^a$ is the vector
field of the {\em inner\/} unit normal to the brane. The quantity $K = K_{ab}
h^{ab}$ is the trace of the symmetric tensor of extrinsic curvature $K_{ab} =
h^c{}_a \nabla_c n_b$ of the brane. The parameter $\epsilon = 1$ if the
signature of the bulk space is Lorentzian, so that the extra dimension is
spacelike, and $\epsilon = -1$ if the signature is $(-,-,+,+,+)$, so that the
extra dimension is timelike. Since the induced metric on the brane is assumed
to be Lorentzian, the signature of the extra dimension coincides with the type
of the unit normal $n^a$ to the brane.  The symbol $L (h_{ab}, \phi)$ denotes
the Lagrangian density of the four-dimensional matter fields $\phi$ the
dynamics of which is restricted to the brane so that they interact only with
the induced metric $h_{ab}$. All integrations over the bulk and brane are taken
with the natural volume elements $\sqrt{- \epsilon g}\, d^5 x$ and $\sqrt{-
h}\, d^4 x$, respectively, where $g$ and $h$ are the determinants of the
matrices of components of the metric in the bulk and of the induced metric on
the brane, respectively, in the coordinate basis. The symbols $M$ and $m$
denote, respectively, the five-dimensional and four-dimensional Planck masses,
$\Lambda$ is the five-dimensional cosmological constant, and $\sigma$ is the
brane tension. In this paper, we use the notation and conventions of
\cite{Wald}.

Variation of action (\ref{action}) gives the equation of motion in the
five-dimensional bulk:
\begin{equation} \label{bulk}
{\mathcal G}_{ab} + \Lambda g_{ab} = 0 \, ,
\end{equation}
and on the brane:
\begin{equation} \label{brane}
m^2 G_{ab} + \sigma h_{ab} = \epsilon M^3 S_{ab} + \tau_{ab} \, ,
\end{equation}
where ${\mathcal G}_{ab}$ and $G_{ab}$ are the Einstein's tensors of the
corresponding spaces, $S_{ab} \equiv K_{ab} - K h_{ab}$, and $\tau_{ab}$
denotes the four-dimensional stress--energy tensor of matter on the brane.

Using the Gauss relation on the brane, one obtains the constraint equation
\begin{equation} \label{constraint}
\epsilon \left( R - 2 \Lambda \right) + K_{ab} K^{ab} - K^2  = 0 \, .
\end{equation}
Then, expressing the extrinsic curvature $K_{ab}$ from (\ref{brane}), one
obtains the following scalar equation on the brane \cite{Shtanov}:
\begin{equation} \label{closed} 
\epsilon M^6 \left( R - 2 \Lambda \right) + \left( m^2 G_{ab} + \sigma h_{ab} -
\tau_{ab} \right) \left( m^2 G^{ab} + \sigma h^{ab} - \tau^{ab}
\right) 
- {1 \over 3} \left( m^2 R - 4 \sigma + \tau \right)^2 = 0\, ,
\end{equation}
where $\tau = h^{ab} \tau_{ab}$.

Before proceeding further, we pose to discuss the significance of the timelike
character of the extra dimension. We emphasize that, in the theory under
consideration, the extra dimension is accessible only for the gravitational
field, while all matter fields are constrained to the brane and are propagating
on the background of the induced Lorentzian metric. Thus, the dynamics of the
matter fields on the brane has the standard properties of quantum field theory
in curved Lorentzian spacetime, and the effect of the extra timelike dimension
tells only in the gravitational sector. An important issue demanding further
investigation is related to the tachyonic nature of the Kaluza--Klein
gravitational modes that could in principle lead to violation of causality and
unitarity on the brane. Some discussion of this issue within the context of
braneworld models with more than one time like dimension can be found in
\cite{DGS,CK}.

In passing, we mention that a generalization of the Randall--Sundrum solution
\cite{RS} to the case of arbitrary signature of the extra dimension and
arbitrary Ricci-flat vacuum brane can be written in the form \cite{CK,Shtanov}
\begin{equation}\label{RS}
ds^2 = \epsilon dy^2 + \exp \left(- {\epsilon \sigma \over 3 M^3} y \right)
h_{\alpha\beta} (x) dx^\alpha dx^\beta \, ,
\end{equation}
where $h_{\alpha\beta} (x)$ are the components of the Ricci-flat metric on the
brane, which is situated at $y = 0$, and the bulk coordinate is in the range $y
\ge 0$.  Equation (\ref{RS}) is supplemented by the constraint
$\Lambda_{\mathrm eff} = 0$, where the expression for $\Lambda_{\mathrm eff}$
is given by Eq.~(\ref{lambda}) below.  One can see that, for $\epsilon = -1$
and negative brane tension $\sigma$, the gravity appears to be `localized' at
the brane (it is, perhaps, more appropriate to speak of `deflation' in the case
of timelike extra dimension).

In this letter, we study cosmological implications of this braneworld theory. We
first consider a homogeneous and isotropic cosmological model with the
cosmological time $t$, scale factor $a (t)$, energy density $\rho (t)$, and
pressure $p (t)$.  In this case, by integrating Eq.~(\ref{closed}), one obtains
the following equation \cite{Shtanov}:
\begin{equation} \label{cosmo}
m^4 \left( H^2 + {\kappa \over a^2} - {\rho + \sigma \over 3 m^2} \right)^2 =
\epsilon M^6  \left(H^2 + {\kappa \over a^2} - {\Lambda \over 6} - {C \over
a^4} \right) \, ,
\end{equation}
where  $C$ is the integration constant, $H \equiv \dot a/a$, and $\kappa = 0,
\pm 1$ corresponds to the spatial curvature of the universe. For $m = 0$, which
corresponds to the Randall--Sundrum limit \cite{RS}, this equation reduces to
\begin{equation}\label{cosmolim}
H^2 + {\kappa \over a^2} = {\Lambda \over 6} + {\epsilon \sigma^2 \over 9 M^6}
+ {2 \epsilon \sigma \rho \over M^6} + {\epsilon \rho^2 \over M^6} + {C \over
a^4} \, .
\end{equation}
Introducing the notation
\begin{equation}\label{newton}
G_{\mathrm N} = {3 \epsilon \sigma \over 4 \pi M^6} \, ,
\end{equation}
\begin{equation}\label{lambda}
\Lambda_{\rm eff} = {\Lambda \over 2} + {\epsilon \sigma^2 \over 3 M^6} \, ,
\end{equation}
one can write Eq.~(\ref{cosmolim}) as follows:
\begin{equation}\label{cosmo1}
H^2 + {\kappa \over a^2} = {\Lambda_{\mathrm eff} \over 3} + {8 \pi G_{\mathrm
N} \rho \over 3} + {\epsilon \rho^2 \over M^6} + {C \over a^4} \, .
\end{equation}
Thus, $G_{\mathrm N}$ is the effective gravitational constant, and
$\Lambda_{\mathrm eff}$ is the effective cosmological constant on the brane.
(For $\epsilon = 1$, Eq.~(\ref{cosmo1}) reduces to the well-known results
\cite{BDEL,cline}.)

A FRW universe described by (\ref{cosmo1}) is embedded in the bulk with the
metric
\begin{equation}\label{metric}
ds^2 = {}- f (r) du^2 + {\epsilon dr^2 \over f (r)} + r^2 d\Omega_\kappa \, ,
\end{equation}
where
\begin{equation}
f (r) = \epsilon \left( \kappa - {\Lambda r^2 \over 6} - {C \over r^2} \right)
\end{equation}
and $d\Omega_\kappa$ is the standard Euclidean three-dimensional metric
corresponding to $\kappa$, which satisfies the vacuum bulk equation
(\ref{bulk}). The position of the brane is described by the equation $r = a
(u)$.

The function $f(r)$ is assumed to be positive in the domain of action of the
coordinate $r$, which, in particular, implies that at least one of the
constants $\kappa$, $\Lambda$ or $C$ must be nonzero. Metric (\ref{metric}) has
singularities and horizons in the bulk for certain values of the constants. We
will not concern ourselves with this issue, assuming that such singularities
can be avoided, for instance, by the introduction of another brane. Although
the presence of
another brane can affect the evolution and spectrum of perturbations (as
in the case in the Randall--Sundrum model), it does not modify the
general equations of the cosmological evolution of the brane.

We first discuss some interesting consequences of Eq.~(\ref{cosmolim}) or
(\ref{cosmo1}) for $\epsilon = -1$. Specifically, as we briefly noted in
\cite{SS}, the fact that the $\rho^2$ term on the right-hand side of
(\ref{cosmo1}) has a negative sign, leads a contracting universe to bounce
instead of reaching the singular state $\rho \to \infty$, $R_{abcd}R^{abcd} \to
\infty$ --- typical of general relativistic `big crunch' singularities. In
order that the subsequent evolution of the universe be compatible with
observations, one requires the brane tension  $\sigma$ to be negative so that
the effective gravitational constant, given by (\ref{newton}), is positive. The
additional requirement that the bounce take place at densities greater than
that during cosmological nucleosynthesis leads to the constraint \cite{cline}
$\vert \sigma\vert \gsim (1{\rm MeV})^4$. Since $\epsilon = -1$, a braneworld
with $\Lambda_{\rm eff} \geq 0$ in (\ref{lambda}), must have a positive bulk
cosmological constant $\Lambda > 0$.

As the universe collapses to high densities, the negative $\rho^2$ term grows
much faster than the remaining terms on the right-hand side of (\ref{cosmo1})
leading to $H \to 0$ and to the inevitable bounce of the braneworld. This
feature appears to be quite generic, since it depends neither upon the equation
of state of matter nor upon the spatial curvature of the universe. The simplest
singularity-free bouncing braneworld model (with $C$, $\kappa$, and
$\Lambda_{\rm eff}$ equal to zero) has the form
\begin{equation}
H^2 = {8 \pi G_{\mathrm N} \rho \over 3} - {\rho^2 \over M^6} \, .
\end{equation}

\begin{figure}[tbh!]
\centerline{ \psfig{figure=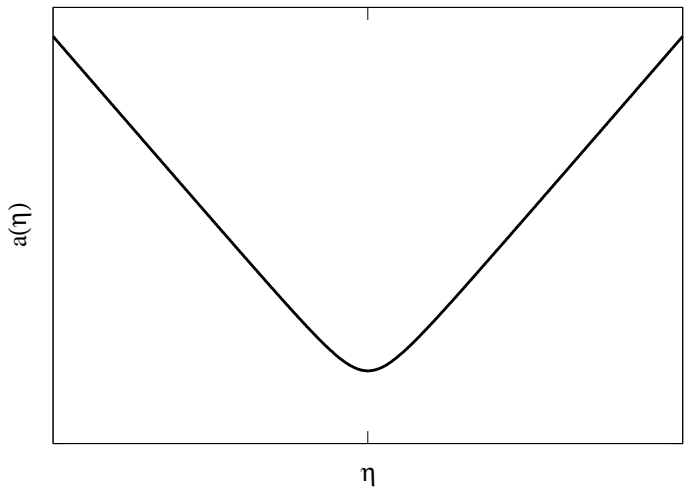,width=0.7\textwidth,angle=-90} }
\bigskip
\caption{\small A bouncing radiation-dominated braneworld ($\eta = \int dt/a$
is the conformal time).} \label{fig:bounce}
\end{figure}

An example of a bouncing universe containing radiation is shown in
Fig.~\ref{fig:bounce}. It is easy to show that a spatially closed universe
($\kappa = 1$) with matter satisfying $\rho + 3p > 0$ will be `cyclic' in the
sense that it will pass through an infinite number of nonsingular
expanding-collapsing epochs. As demonstrated in \cite{kanekar}, a massive
scalar field in such a universe usually leads to an increase in the amplitude
of consecutive expansion cycles and to a gradual amelioration of the flatness
problem.

In passing, we note that cosmological braneworld equations with $\rho^2$
correction terms of negative sign on the right-hand side were also recently
discussed in \cite{CV} in the context of the Randall--Sundrum two-brane model
with a bulk scalar field stabilizing the radion.  We also note that bouncing
and cyclic braneworlds were discussed in \cite{MP} in the context of the usual
theory with spacelike extra dimension where the bulk metric is that of a
charged anti de~Sitter black hole.  In this case, the bounce occurs if the
black-hole charge is sufficiently high.  In our model, the bounce is quite a
generic feature.

Next, let us consider the case of a homogeneous but anisotropic braneworld
described by the Bianchi~I metric
\begin{equation} \label{metric-i}
ds^2 = - dt^2 + \sum_{i=1}^3 a_i^2 (t) \left( dx^i \right)^2 \, .
\end{equation}
In this case, Eq.~(\ref{closed}) with $m = 0$ gives the following closed
equation on the brane:
\begin{equation} \label{cosmo-i}
6 \dot H + 12 H^2 + \sum_{i=1}^3 \left( H_i - H \right)^2 = 4 \Lambda_{\mathrm
eff} + {2\epsilon \over 3 M^6} \left[ \sigma (\rho - 3 p) - \rho (\rho + 3 p)
\right] \, ,
\end{equation}
where we used the notation of (\ref{lambda}) and
\begin{equation}
H_i = {\dot a_i \over a_i}\, , \quad H = \frac13 \sum_{i=1}^3 H_i = {\dot a
\over a}  \, , \quad a = \left( a_1 a_2 a_3 \right)^{1/3} \, .
\end{equation}

The last term on the left-hand side of Eq.~(\ref{cosmo-i}) is the shear scalar
of the spatial section of the universe: $\sum\limits_{i=1}^3 \left( H_i - H
\right)^2 \equiv \sigma_{\alpha\beta} \sigma^{\alpha\beta}$.  Because of the
presence of this term, and because the evolution of the shear tensor
$\sigma_{\alpha\beta}$ is not specified on the brane (see \cite{MSS}),
Eq.~(\ref{cosmo-i}) cannot be integrated completely, but the result of its
integration can be written in the form
\begin{equation} \label{cosmo-ii}
H^2 + {1 \over 3 a^4} \int \sum_{i=1}^3 \left( H_i - H \right)^2 a^3 \dot a \,
dt = {\Lambda_{\mathrm eff} \over 3} + {8 \pi G_{\mathrm N} \rho \over 3} +
{\epsilon \rho^2 \over M^6} + {C \over a^4} \, ,
\end{equation}
where the notation is the same as in (\ref{newton}) and (\ref{lambda}).  We
emphasize that Eq.~(\ref{cosmo-ii}) describing the evolution of the Bianchi~I
brane is absolutely general.

As we noted, the behaviour of the second term on the left-hand side is not
specified and, in principle, can be arbitrary. Nevertheless, our conclusion
about the bounce of the contracting universe in the case of a timelike extra
dimension ($\epsilon = - 1$) remains valid as long as the shear scalar
$\sigma_{\alpha\beta} \sigma^{\alpha\beta} \equiv \sum\limits_{i=1}^3 \left(
H_i - H \right)^2$ does not grow faster than $a^{-8}$ as $a \to 0$ during the
contraction of a radiation dominated universe. As before, the bounce is caused
by the negative $\rho^2$ term on the right-hand side. However, in the case of
the anisotropic model, the bounce is taking place only as regards the overall
expansion $\dot a / a$, while the behaviour of the shear tensor remains
unspecified in general; this leaves open the possibility that such a universe
bounces only in a given spatial patch.

As an example of bouncing behaviour, the shear can be specified \cite{MSS,CS}
by the additional assumption that the so-called nonlocal energy density on the
brane ${\cal U}$, which is determined by the projection of the Weyl tensor to
the brane, either vanishes or is negligible. This leads to \cite{MSS}
\begin{equation}
\sigma_{\alpha\beta} \sigma^{\alpha\beta} \equiv \sum_{i=1}^3 \left( H_i - H
\right)^2 = {6 \Sigma^2 \over a^6} \, , \quad \Sigma = {\mathrm const} \, ,
\end{equation}
which clearly reinforces our earlier conclusion about the bounce. Indeed,
the integral in Eq.~(\ref{cosmo-ii}) can now be evaluated to give
\begin{equation} \label{cosmo-iii}
H^2 = {\Lambda_{\mathrm eff} \over 3} + {8 \pi G_{\mathrm N} \rho \over 3} +
{\epsilon \rho^2 \over M^6} + {C \over a^4} + {\Sigma^2 \over a^6} \, ,
\end{equation}
and it is immediately clear that the $\rho^2$ term on the right-hand side,
which grows as $a^{-8}$ during the radiation-dominated contraction stage,
dominates over the last (shear) term. The same result is obtained in a slightly
more general setup \cite{Toporensky}, if the nonlocal energy density ${\cal U}$
is assumed to behave like radiation, ${\cal U} \propto a^{-4}$.  Note that,
although the universe bounces,  the behaviour of the individual components of
the shear tensor $\sigma_{\alpha\beta}$ (hence also of the individual
components $H_i$, $i = 1,2,3$) is not fully specified in the above examples.

The collapse of a scalar-field dominated universe is even more likely to lead
to a bounce because the energy density $\rho_\phi$ of the field $\phi$ during
contraction becomes kinetic-dominated, $\dot \phi^2 /2 \gg V(\phi)$, so that
$\rho_\phi \propto a^{-6}$, which makes the $\rho^2$ term in (\ref{cosmo1})
grow much faster than in the case of radiation \cite{kanekar}.

So far, we have been assuming $m = 0$ in the action (\ref{action}). Let us
return to a homogeneous and isotropic universe but drop this constraint, i.e.,
extend our study to braneworld models whose action contains the induced
curvature term on the brane. The cosmological equation (\ref{cosmo})
corresponding to this model can be solved with respect to the Hubble parameter:
\begin{eqnarray}
H^2 + {\kappa \over a^2} &=& {\rho + \sigma \over 3 m^2} + {2 \epsilon \over
\ell^2} \left[1 \pm \sqrt{1 + \epsilon \ell^2 \left({\rho + \sigma \over 3 m^2}
- {\Lambda \over 6} - {C \over a^4} \right)} \right] \nonumber \\
&=& {\Lambda \over 6} + {C \over a^4} + {\epsilon \over \ell^2} \left[1 \pm
\sqrt{1 + \epsilon \ell^2 \left({\rho + \sigma \over 3 m^2} - {\Lambda \over 6}
- {C \over a^4} \right)} \right]^2 \, , \label{solution}
\end{eqnarray}
where we introduce the length parameter
\begin{equation}
\ell = {2 m^2 \over M^3} \, .
\end{equation}
Similar equations were obtained in \cite{CHD} for the case of $\epsilon = 1$.
The `$\pm$'signs in (\ref{solution}) correspond to two different ways of
bounding the bulk space by the brane, depending on whether the inner normal to
the brane points in the direction of increasing or decreasing bulk coordinate
$r$ in (\ref{metric}). Alternatively, the two different signs in
(\ref{solution}) could correspond to the two possible signs of the
five-dimensional Planck mass $M$. The  model with the `$-$' sign is the one
that passes smoothly to the previously considered case of $m = 0$ in the sense
that the right-hand side of (\ref{solution}) with `$-$' sign can formally be
expanded in powers of $m$.  The right-hand side of (\ref{solution}) with `$+$'
sign formally diverges as $m \to 0$.

Since model (\ref{solution}) with `$-$' sign passes smoothly to the model
described by Eq.~(\ref{cosmo1}) as $m \to 0$, it is clear that, for
sufficiently small values of $m$, the evolution described by (\ref{solution})
with `$-$' sign will not be very different from that described by the limiting
Eq.~(\ref{cosmo1}) with the same initial values of the scale factor and energy
density, and the bounce will take place in the case of $\epsilon = -1$. This is
explained by the fact that the last term in the second line of (\ref{solution})
has a negative sign and grows by absolute value during the contracting phase.
In particular, the bounce will definitely take place in the case of spatially
flat or closed universe with zero dark-energy term ($C = 0$) if $\Lambda \ell^2
/ 6 < 1$ $\Rightarrow$ $\Lambda m^4 / M^6 < 3/2$.  If the value of $m$ is not
so small as to lead to the bounce of the contracting universe, then such a
universe will reach a singularity similar to that of the case $\epsilon = 1$
described in our paper \cite{singular}. Specifically, as the universe
collapses, its energy density $\rho$ grows and can, under certain assumptions,
exceed the dark radiation term under the square root in (\ref{solution}).
Further increase in $\rho$ will cause the expression under the square root to
reach zero, heralding the formation of a cosmological singularity beyond which
the solution cannot be continued. This singularity is unusual, since the energy
density and pressure remain finite, while the space-time curvature of the brane
and its extrinsic curvature tend to infinity as the singularity is approached
(see \cite{singular} for more details).

Our general conclusions are therefore the following:

1. The four-dimensional cosmological evolution of a Randall--Sundrum-type model
with a timelike extra dimension is non-singular both in the past and in the
future provided matter satisfies the weak energy condition $\rho + p \geq 0$.
The `big crunch/big bang' singularities are also absent in a broad class of
anisotropic Bianchi~I braneworld models with a timelike extra dimension.

2. The presence of an `induced' brane curvature term in the higher-dimensional
action (\ref{action}) can trigger the formation of an initial or final
singularity even when the extra dimension is timelike. However, the properties
of this singularity are unusual since the Hubble parameter remains finite while
the curvature tensor diverges as the singularity is approached.

Thus, braneworlds with timelike extra dimensions have properties which are
fundamentally different both from standard general-relativistic behaviour and
from the properties of braneworld models with spacelike extra dimensions, and
which could have interesting cosmological implications.  The important issue of
the tachyonic nature of Kaluza--Klein gravitational modes in such theories and
the related problem of unitarity has been discussed in \cite{DGS,CK} but
demands more extensive examination.

\bigskip

{\em Acknowledgments\/}: The authors acknowledge support from the
Indo-Ukrainian program of cooperation in science and technology sponsored by
the Department of Science and Technology of India and Ministry of Education and
Science of Ukraine. The work of Yu.~S.\@ was also supported in part by the
INTAS grant for project No.~2000-334. The authors acknowledge useful
discussions with Katrin Heitmann and Salman Habib.

\end{document}